\documentclass[11pt,twoside]{article}

\usepackage{asp2004}
\usepackage{epsf}
\usepackage{psfig}
\usepackage{lscape} 

\def\lsi{\raise0.3ex\hbox{$<$\kern-0.75em\raise-1.1ex\hbox{$\sim$}}}
\def\gsi{\raise0.3ex\hbox{$>$\kern-0.75em\raise-1.1ex\hbox{$\sim$}}}
\newcommand{\lsim}{\mathop{\lsi}}

\newcommand{\src}{\mbox{\footnotesize{src}}}
\newcommand{\obs}{\mbox{\footnotesize{obs}}}

\begin{document}
\title{Constraining Gravitational Theories by Observing Magnetic White Dwarfs}
\author{Oliver Preuss$^1$, Stefan Jordan$^{2}$, Mark P. Haugan$^3$,
and Sami K. Solanki$^1$}

\affil{$1$ Max-Planck-Institut f\"ur Sonnensystemforschung,
             D-37191 Katlenburg-Lindau, Germany \\
       $2$ Astronomisches Rechen-Institut, D-69120 Heidelberg, Germany\\    
       $3$ Department of Physics, Purdue University 1396, West Lafayette, Indiana 47907, USA}

\begin{abstract}
  Under the assumption of a specific nonminimal coupling of torsion 
  to electromagnetism, spacetime is birefringent in the presence of a 
  gravitational field leading to depolarization of light emitted from extended 
  astrophysical sources. We use polarimetric data of the magnetic white 
  dwarf $\mbox{RE J0317-853}$ to set for the very first time constraints on 
  the essential coupling constant for this effect, giving $k^2 \lsim\, 
  (22 \,\mbox{m})^2 $.  
\end{abstract}

\section{Introduction}
  Efforts to develop a quantum theory of gravity or a complete, 
  consistent and unified theory of matter and all its interactions are 
  rich and compelling sources of speculation about new physics 
  beyond the scope of the standard model of particle physics or of general
  relativity.
  The effective field theories that emerge as low-energy limits of string 
  theories are littered with new fields and interactions.  Since cherished 
  symmetries like CPT and Lorentz invariance can be broken in 
  these contexts (Colladay et al. 1998), high-precision experimental and 
  observational tests of these symmetries offer particularly effective ways 
  of searching for evidence of new physics.   
  This project was motivated by current speculations assuming a special 
  nonminimal coupling of torsion to electromagnetism (Preuss et al. 2004,
  Solanki et al. 2004). Members of the class of metric-affine theories 
  (Hehl et al. 1995) of gravity feature torsion and/or nonmetricity 
  gravitational fields in addition to a symmetric second-rank tensor 
  gravitational potential.  
  We suggest, in addition to the conventional Maxwell Lagrangian, the 
  additional nonminimal coupling
  \begin{equation}
     L_{EM} = k^2 *(T_\alpha \wedge F)\,T^\alpha \wedge F \quad , 
  \end{equation}
  see (Itin \&\ Hehl 2003) for other possibilities, where $k$ is a coupling 
  constant with the dimension of length, $*$ denotes the Hodge dual, $T$ 
  denotes the torsion and $F$ the electromagnetic field, which 
  is related in the usual way to its potential $A$. This addition is 
  consistent with gauge invariance and, so, with charge conservation. 

  We use static, spherically symmetric torsion fields (Tresguerres, 
  1995) in which a nonminimal coupling to electromagnetism singles out linear polarizations
  with a fractional difference in their propagation speeds (Haugan \&\ Kauffmann 1995).
  We are interested in the effect that this differential propagation has on 
  the polarization of light as it travels from a localized source on the 
  stellar surface to an observer. This is determined by the phase shift $\Delta \Phi$ 
  that accumulates between the polarization components singled out by the 
  stellar field as the light propagates.We find in this case 
  \begin{equation}\label{dphi}
    \Delta \Phi = \sqrt{2 \over 3} {{2 \pi k^2 M} \over {\lambda R^2}} 
    {{(\mu +2) (\mu - 1)} \over {\mu + 1}}, 
  \end{equation}
  where $\mu$ denotes the cosine of the angle between the line of sight
  and the normal on the stellar surface ($\mu=1$: stellar disk center, 
  $\mu=0$: limb), $\lambda$ is the light's wavelength, $R$ is the stellar 
  radius and $M$ the stellar mass in geometrized units.  

\begin{figure}[t]
\begin{minipage}[t]{0.60\textwidth}
\centerline{\psfig{figure=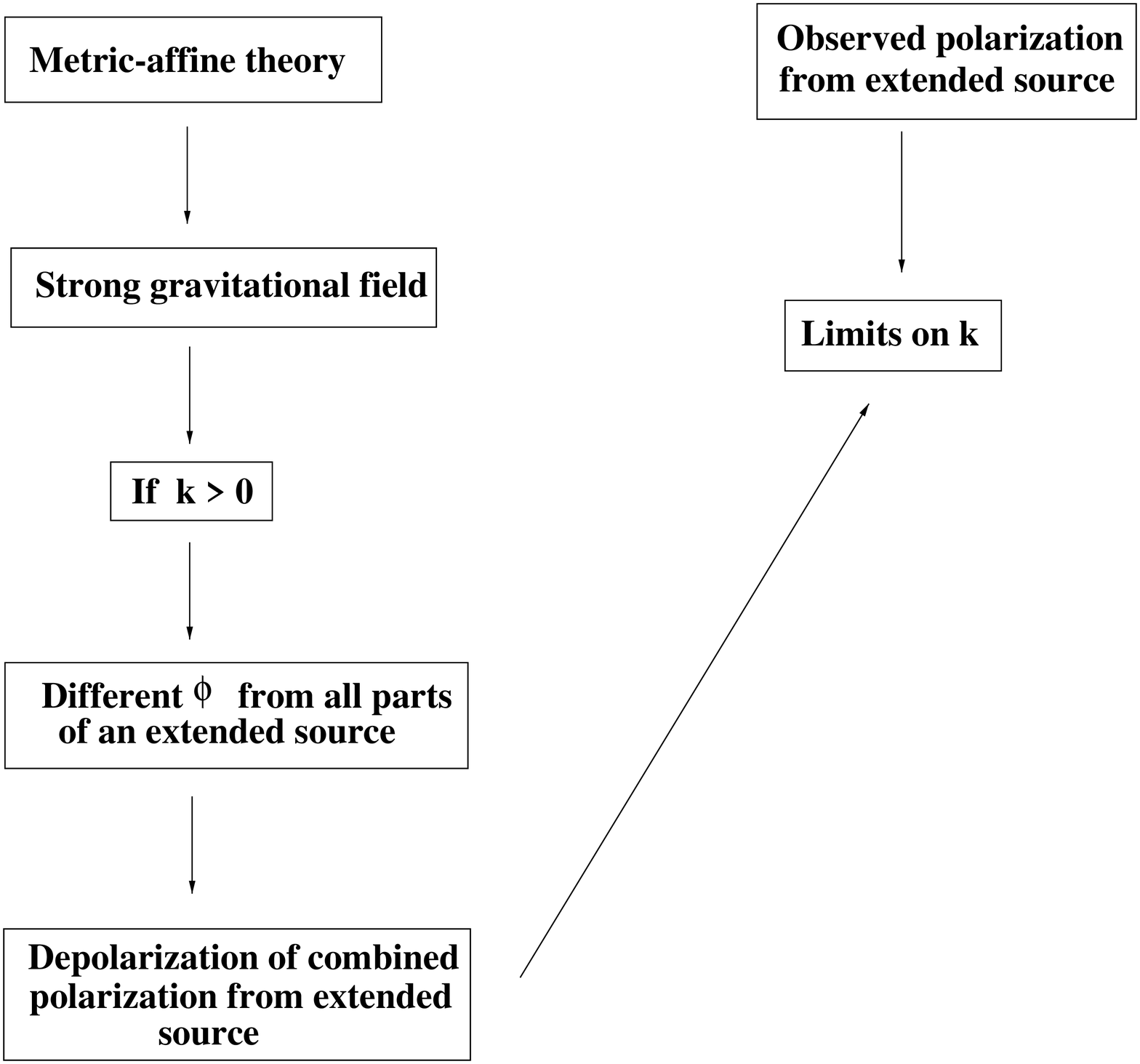,width=\textwidth}}
      \caption{General scheme for setting limits on gravity-induced birefringence}
\end{minipage}
\hfill
\begin{minipage}[t]{0.39\textwidth}
\centerline{\psfig{figure=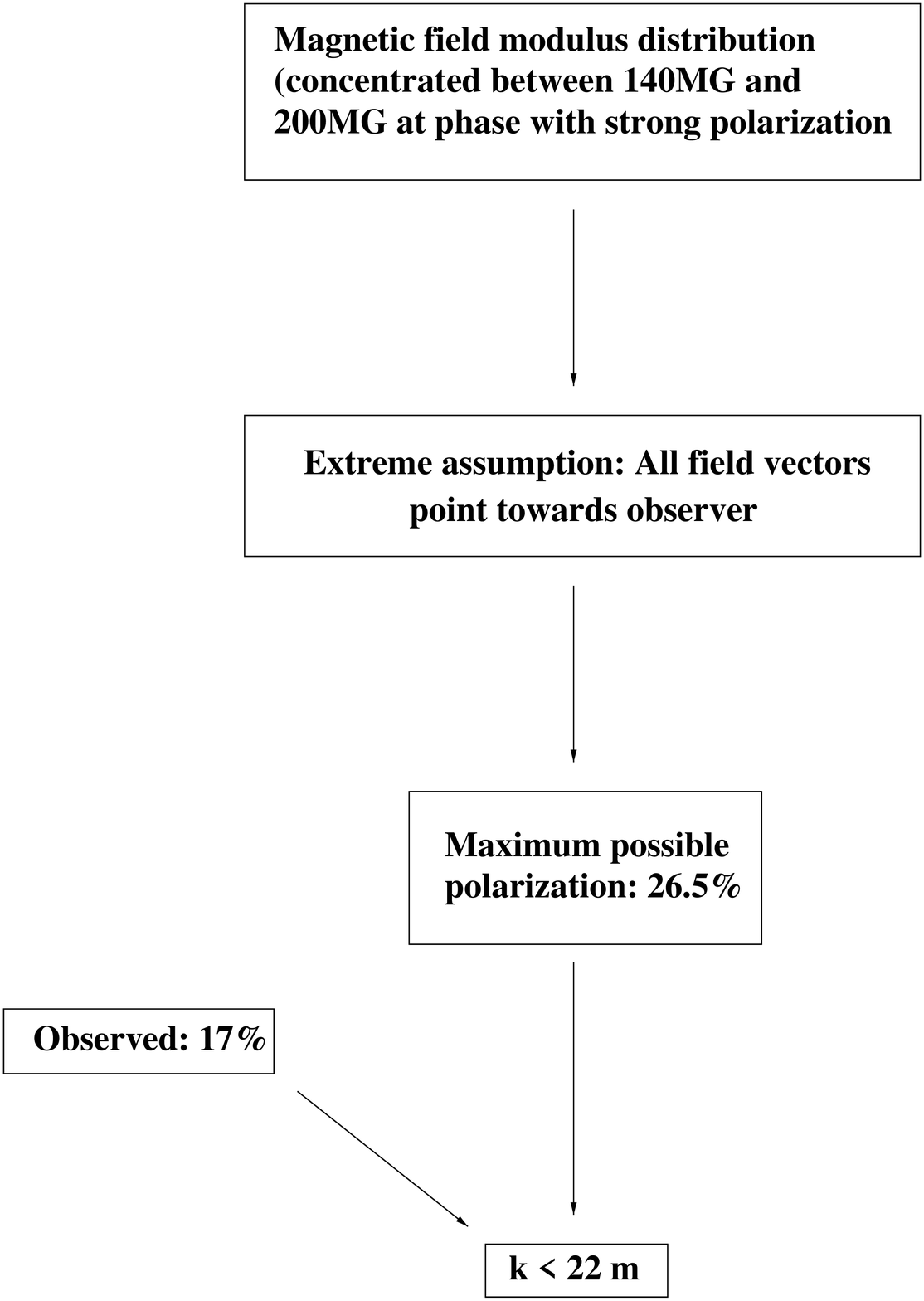,width=\textwidth}}
      \caption{Scheme for our specific case}
\end{minipage}
\end{figure}
  
\section{Data, analysis and results}
    If an extended source covering a range of $\mu$ values is observed 
    then light emitted from different points suffers different phase 
    shifts (\ref{dphi}) and, so, adds up to an incoherent superposition. 
    This yields a reduction of the observed polarization relative to the 
    light emitted from the source:
    $(U_{\obs}^2 + V_{\obs}^2)^{1/2} < (U_{\src}^2 + V_{\src}^2)^{1/2}$. 
    Since the rotationally modulated polarization from magnetic white 
    dwarfs can only be produced by an extended source any 
    observed (i.e. non-zero) degree of polarization provides a limit 
    on the strength of birefringence induced by the star's gravitational 
    field. 

    The polarized radiation from white dwarfs is produced by the magnetic
    field. Since the disk of a white dwarf is unresolved, only the total 
    polarization from all surface elements is observable:
    \begin{equation}\label{vl}
      V_{\lambda, {\rm tot}}(k^2) = 2\pi\int\int \,V_{\lambda}(\mu,B,B_{\|})\,\cos(\Delta\Phi)
      \, \mu \, d\theta\, d\phi \,\, , 
    \end{equation}         
   
    \noindent where the Stokes parameter  $V_{\lambda}$ changes over the
    visible hemisphere and depends on the wavelength $\lambda$, the location $\mu$ 
    (limb darkening), the total magnetic field strength $B(\theta,\phi)$, the 
    line-of-sight component $B_{\|}(\theta,\phi)$, and on the parameters of the 
    stellar atmosphere. Gravitational birefringence reduces the polarization by
    means of $\cos(\Delta\Phi)$, see Fig.~1. 
    
    The Stokes parameters can be calculated by solving the radiative transfer 
    equations through a magnetized stellar atmosphere on a large number of surface 
    elements on the visible hemisphere. If the star is rotating, the spectrum and 
    polarization pattern changes according to the respective magnetic field distribution 
    visible at a particular moment. To obtain the degree of circular polarization, 
    Eq.~(\ref{vl}) has to be divided by the total stellar flux $I_{\lambda, {\rm tot}}$ 
    emitted to the observer at wavelength $\lambda$. 
   
    The rapidly rotating hot magnetic white dwarf $\mbox{RE J0317-853}$ (Barstow et al. 1995) 
    is best suited for setting limits on gravitational birefringence, since it has 
    a strong gravitaitional field ($1.35\,M_{\odot}, 0.0035\, R_{\odot}$), and, 
    with $V_{\lambda, {\rm obs}}/I_{\lambda,{\rm tot}}$ of  $17$\%  at $5760$\AA\ the 
    the highest known level of circular polarization in a white dwarf (Jordan \&\ 
    Burleigh 1999). We use a value of 17\,\% here instead of 20\,\%  as in Preuss et al. 
    (2004) since this more conservatively takes into account that higher value may only be reached 
    by some noise peaks. The analysis of time resolved HST flux spectra in the UV
    has shown that the distribution of the field moduli is approximately that of an 
    off-centered magnetic dipole oriented obliquely to the rotation axis with a polar 
    field strength of $B_d=363$\,MG, leading to visible surface field strengths 
    between $140$ and $730$\,MG (Burleigh et al. 1999). This model can also approximately 
    fit the optical spectra (Jordan et al. in prep.), which means that the 
    distribution of the magnetic field moduli - but not necessarily of the longitudinal 
    components -  is correctly described. This result is completely independent of the 
    magnitude of the gravitational birefringence. 
    From radiative transfer calculations it follows that at the phase of rotation
    when the maximum value of 17\%\ polarization at 5760\,\AA\ is measured, almost the 
    entire visible stellar surface is covered by magnetic fields between $140$
    and $200$\,MG, with only a small tail extending to maximum field strengths of 
    $530$\,MG. Using this special field geometry 
    we calculated a histogram distribution of the visible surface magnetic field 
    strengths in order to set sharp limits on gravitational birefringence.     
    For each field strength bin of the histogram we calculated the maximum circular
    polarization from radiative transfer calculations by assuming that the field vector 
    always points towards the observer. The total maximum polarization from the whole 
    visible stellar disk without gravitational birefringence is then calculated by 
    adding up the contributions from each field strength bin weighted with its relative 
    frequency. This results in $V_{\lambda, {\rm max}}/I_{\lambda,{\rm tot}} = 26.5$\%. 

    Assuming that the reduction to the observed 
    $V_{\lambda, {\rm obs}}/I_{\lambda,{\rm tot}}= 17$\% 
    is entirely due to gravity-induced depolarization -- and not due to the fact that 
    in reality not all field vectors point towards the observer -- we can calculate
    an upper limit for this effect of $k^2 \lsim (22\,{\rm m})^2$, see Fig.~2. Since there is 
    always a small uncertainty in determing the exact mass of a white dwarf, we also 
    calculated an upper limit on $k^2$ assuming a lower mass of $1\,M_{\odot}$. 
    This leads to $k^2 \lsim (25.5\,{\rm m})^2$.    

    An even more extreme assumption would be to use the maximum circular polarization 
    predicted by all field strengths in the interval $140-530$\,MG (reached at
    $530$\,MG) for the whole stellar disk. Then we obtain $V_{\lambda, {\rm max}}/I_{\lambda, 
    {\rm tot}}=48.3$\%\ and an upper limit of $k^2 \lsim (33\,{\rm m})^2$.
    Independent from any dipole model and without any reference to radiative transfer calculations 
    the assumption of 100\%\ emerging polarization leads to $k^2 \lsim (47.5\,{\rm m})^2$.
    In order to compare the quality of our method with previous similar results (Solanki 
    et al. 1999)  we also set new upper limits on the NGT parameter $\ell^2_{\star}$ wich 
    also causes gravitational birefringence in case of $\ell^2_{\star}\neq 0$. Assuming $V_{\lambda, 
    {\rm max}}/I_{\lambda,{\rm tot}} = 26.5$\% leads to $\ell^2_{\star} \lsim (1.8\,{\rm km})^2$
    in contrast to $\ell^2_{\star} \lsim (4.9\,{\rm km})^2$, determined for the white dwarf
    GRW $+70^{\circ}8247$.    
\section{Conclusions}
  The spectropolarimetric observations of the massive $\mbox{RE J0317-853}$ 
  impose new strong constraints on the birefringence of space-time in the 
  presence of a gravitational field with an upper limit for the relevant 
  coupling constant $k^2$ of $(22\,\mbox{m})^2$ -- or $(47.5\,\mbox{m})^2$ 
  for the most conservative assumptions. 
  Since gravity-induced birefringence violates the Einstein equivalence 
  principle, our analysis also provides a test of this foundation of general 
  relativity and other metric theories of gravity. We consider as a specific 
  case a metric-affine theory that couples the electromagnetic field nonminimally 
  with torsion and for which a static spherically symmetric solutions has been found.
  Considerably tighter limits based on the same astronomical source could be 
  provided by measurements of circular polarization in the FUV (in particular 
  associated with Ly$\alpha$ absorption features) and also by a consistent 
  model for the magnetic field geometry which reproduces the spectropolarimetry 
  measurements in the optical.
  
 \acknowledgements{ We are grateful to D.T. Wickramasinghe and F.W. Hehl
 for helpful remarks and valuable discussions. This research has made use of 
 NASA's ADS Abstract Service.}

\end{document}